\documentclass{ws-ijbc}
\usepackage{ws-rotating} 
\begin{document}

\catchline{}{}{}{}{} 

\markboth{Katsanikas et al}{Chains of tori and filaments in a 4D surface of section}

\title{ Chains of rotational tori and filamentary structures close to high
  multiplicity periodic orbits in a 3D galactic potential}
\author{M. KATSANIKAS}
\address{Research Center for Astronomy, Academy of Athens\\
  Soranou Efessiou 4,  GR-11527 Athens, Greece}
\address{Section of Astrophysics, Astronomy and Mechanics, \\Department of
  Physics, University of Athens, Greece\\ mkatsan@academyofathens.gr}
\author{P.A. PATSIS}
\address{Research Center for Astronomy, Academy of Athens\\
  Soranou Efessiou 4,  GR-11527 Athens, Greece\\patsis@academyofathens.gr}
\author{A.D. PINOTSIS}
\address{Section of Astrophysics, Astronomy and Mechanics, \\Department of
  Physics, University of Athens, Greece\\ apinots@phys.uoa.gr}

\maketitle

\begin{history}
\received{(to be inserted by publisher)}
\end{history}

\begin{abstract}
This paper discusses  phase space structures  encountered  in the neighborhood 
of periodic orbits with high order multiplicity in a 3D autonomous Hamiltonian 
system with a potential of galactic type. We consider 4D spaces of section and 
we use the method of color and rotation [Patsis and Zachilas 1994] in order to
visualize  them. As examples we use the  case of two orbits, one 2-periodic
and one 7-periodic. We investigate the structure of 
multiple tori around them in the 4D surface of section and in  addition  we  
study the  orbital behavior in the  neighborhood of the  corresponding simple 
unstable  periodic orbits. By considering initially a few  consequents in the  
neighborhood of the orbits in both cases we find a  structure in the space of 
section, which is in direct correspondence with  what  is observed in a 
resonance zone  of  a 2D autonomous Hamiltonian system. However, in our 3D 
case we have instead of  stability islands  rotational tori, while the  
chaotic zone connecting the points of the unstable periodic orbit is replaced 
by filaments extending in 4D following a smooth color variation. For more 
intersections, the consequents of the orbit which started in the neighborhood 
of the unstable periodic orbit, diffuse in phase space and form a cloud that 
occupies a large volume surrounding the region containing the rotational
tori. In  this cloud the colors of the points are mixed. The same structures 
have been observed in the neighborhood of all m-periodic orbits we have 
examined in the system. This indicates a generic behavior.
\end{abstract}

\twocolumn{

\section{Introduction}  

Recently, Katsanikas and Patsis [2011] (hereafter KP11) studied the 
structure of the phase space in the neighborhood of simple periodic orbits in 
a 3D autonomous Hamiltonian system of galactic type. In the present paper we 
extend this work and we investigate the orbital behavior in the neighborhood 
of periodic orbits with high order multiplicity. Especially, we study the 
orbital  behavior in the neighborhood of a stable 2-periodic and a stable 
7-periodic orbit and in the neighborhood of the accompanying simple unstable  
periodic orbits. 
\par In Cartesian coordinates $(x,\dot x,y,\dot y,z,\dot z)$, if we consider 
our surface of section to be  defined by $y=0$, a ``2-periodic'' orbit is one 
that closes after 2 intersections with the surface of section  when 
$\dot{y}>0$. This is a periodic orbit of multiplicity 2. In the same way, in 
general, an ``$m$-periodic'' orbit is a periodic orbit of multiplicity $m$.
Here we study two cases with $m=2$ and $7$. For the visualization  of the 4D 
surfaces  of section  we use the method  of color and rotation
[Patsis and Zachilas 1994]. With this method we plot the  consequents in a 
3D subspace of the 4D  surface  of section and every  consequent is  colored  
according to its location in the 4th dimension. For a  description of this 
method, the meaning  of viewing angles etc., see KP11. 

Our Hamiltonian is of the form
\begin{eqnarray}
H(x, y, z, \dot x, \dot y, \dot z)=
\nonumber\\
\frac {1}{2}(\dot x^2 + \dot y^2 + \dot z^2) + \Phi(x, y, z)
\nonumber\\
- \frac{1}{2} \Omega_b^2 (x^2 + y^2)
\end{eqnarray}

where  $\Phi(x,y,z)$ is the potential we used in our
applications, i.e.:

\begin{eqnarray}
\Phi(x,y,z)= 
\nonumber\\
-\frac{GM_{1}}{(x^2+ \frac{y^2}{q_a^2} +[a_{1}+(\frac{z^2}{q_b^2}+
b_{1}^2)^{1/2}]^2)^{1/2}}-
\nonumber\\
 \frac{GM_{2}}{(x^2+ \frac{y^2}{q_a^2} +[a_{2}+(\frac{z^2}{q_b^2}+b_{2}^2)
^{1/2}]^2)^{1/2}}
\nonumber\\
\end{eqnarray}

This potential is the same  as in KP11, i.e. a 
triaxial double Miyamoto disk rotating around its short axis $z$ with angular  
velocity  $\Omega_b$=60~$km s^{-1} kpc^{-1}$.

In our units, distance $R$=1 corresponds to 1 kpc. The velocity unit corresponds to 209.64~$km/sec$. For the Jacobi constant
Ej=1 corresponds to 43950 $(km/sec)^2$. For the rest of the parameters we have used the
following values: $ a_{1}=0 \; kpc,\; b_{1}=0.495
\; kpc,\; M_{1}=2.05 \times 10^{10} \; M_{\odot},\; a_{2}=7.258 \;
kpc,\; b_{2}=0.520 \; kpc,\; M_{2}=25.47 \times 10^{10} \;
M_{\odot},\; q_a = 1.2,\; q_b = 0.9$. The parameters $q_a, q_b$ determine the geometry
of the disks, while $a, b$ are scaling factors.

The trajectories are calculated numerically by using a 4$^{th}$ order Runge-Kutta scheme, with a constant time step that secures a 13 digits precision. A typical integration for $10^{4}$ consequents with a Core 2 Duo CPU/2.2GHz computer takes about 3.6 minutes of real time.  
 
The calculation  of the linear stability of a  periodic orbit is based 
on the method of Broucke [1969] and Hadjidemetriou [1975]. By this method we
calculate the stability indices b1, b2 and the quantity $\Delta$ (following 
the notation of Contopoulos and Magnenat [1985]). Depending on the values of 
the stability indices and that of $\Delta$, a periodic orbit can be stable 
(S), simple unstable (U), double unstable (DU) and complex unstable 
($\Delta$). For definitions see Contopoulos and Magnenat [1985]. For a 
generalization of this method to systems with higher than 3 degrees of freedom 
see Skokos [2001].

\section{Spaces of Section}
\subsection{Orbits}
The stable and simple unstable  orbits of the 2-periodic case  belong to 
the 3D families of 2-periodic orbits ``$s$'' (an initially stable family) 
and ``$u$'' (an initially simple unstable family). They are bifurcations    
of a planar 3/1 family on the equatorial plane. The morphology of the  orbits of
the families $s$ and $u$ is depicted in Figs.~\ref{2orb1} and ~\ref{2orb2} 
respectively. 
\par We call the 3D 7-periodic families  
$s1$ (an initially stable family) and $u1$ (an initially simple 
unstable family). They  are bifurcations of the family x1v1 [Skokos et al 
  2002a,b] that is associated with the vertical resonance 2/1 in our galactic 
system. In Figs.~\ref{7orb1},~\ref{7orb2} we show  the morphology of the
orbits of the families $s1$ and $u1$ respectively.
\begin{figure*}
\begin{center}
\includegraphics [scale =0.6]{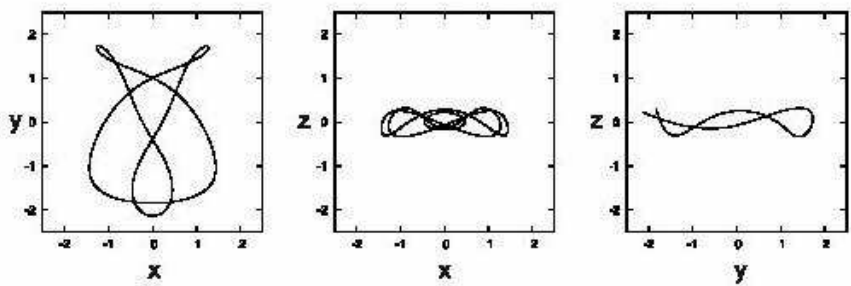}
\caption{ Projections of a stable orbit of the 2-periodic family $s$ at 
Ej= $-4.33035$.}
\label{2orb1}
\end{center}
\end{figure*}

\begin{figure*}
\begin{center}
\includegraphics [scale =0.6]{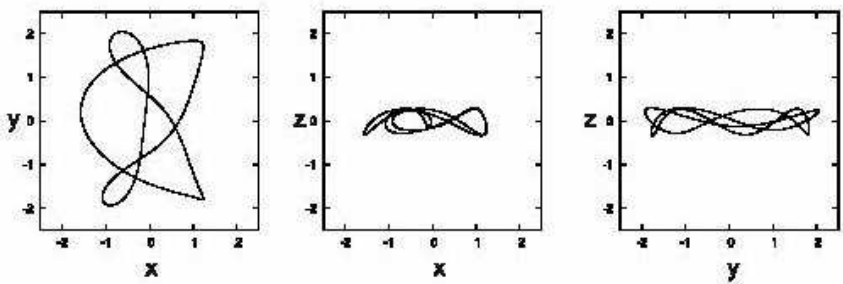}
\caption{Projections of a simple unstable orbit of the 2-periodic family $u$ at 
Ej= $-4.33035$.}
\label{2orb2}
\end{center}
\end{figure*}
\begin{figure*}
\begin{center}
\includegraphics [scale =0.5]{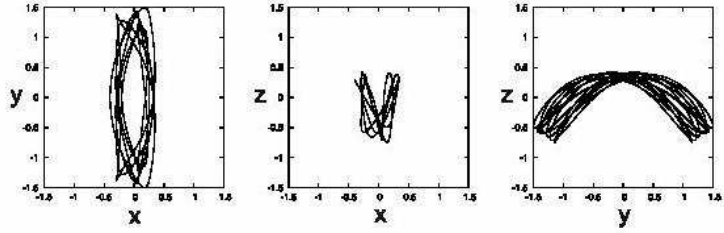}
\caption{Projections of a stable orbit of the 7-periodic family $s1$ at 
Ej= $-4.622377$.}
\label{7orb1}
\end{center}
\end{figure*}

\begin{figure*}
\begin{center}
\includegraphics [scale =0.5]{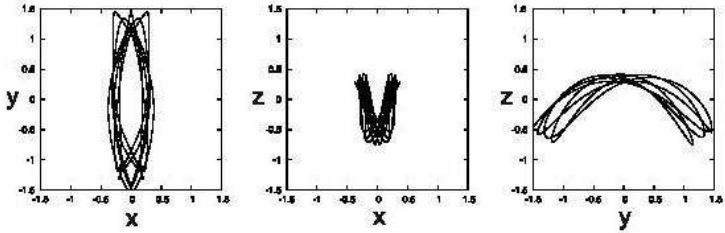}
\caption{Projections of a simple unstable orbit of the 7-periodic family $u1$ at 
Ej= $-4.622377$.}
\label{7orb2}
\end{center}
\end{figure*}

\subsection{4D spaces of section in the neighborhood  of the  
2-periodic orbits}
\subsubsection{Phase space structure close to s}
We consider first the stable 2-periodic orbit $s$ at Ej=$-4.33035$. We trace it at the 
initial conditions $(x_0,z_0,\dot{x_0},\dot{z_0})=(1.0867145,0.2555081,-0.56256937,0.049693274)$.
In the neighborhood of $s$ we observe two tori, when
we perturb its initial condition $x_0$, by $\Delta x$ belonging to the intervals
$10^{-4}\leq\Delta x \leq4\times 10^{-2}$ and  $-10^{-4}\leq\Delta x \leq-5\times 10^{-2}$.
All these perturbations of the initial conditions are isoenergetic displacements on the surfaces of section.
For example if we take $\Delta x=4 \times 10^{-2}$ we observe two tori in 
Fig.~\ref{2tor1} around the two points (black dots) of the stable 
2-periodic orbit. On the two tori we observe a smooth color succession. This means that the succession of the 
colors of the consequents on a given structure follows the succession of the colors on the color bar at the right side of the figure. The consequents are colored according to the value of the coordinate, that is not used in the 3D spatial projection. In all diagrams of this paper the color is given according to the $\dot z$ coordinate. The color values in the color bars are normalized to the [0,1] interval corresponding to [$\dot z_{min}, \dot z_{max}$]. For the details of the method the reader may refer to KP11. The 
smoothness of the color variation can be checked by looking at the color bar, 
on the right of Fig.~\ref{2tor1}, where we see that the color variation is 
between neighboring shades. The colors on the upper left torus vary from red 
to orange to yellow and to green, while the colors on the other torus vary  from  
green to light blue to blue. This means that the fourth value of the 
consequents has a smooth  distribution in the 4th dimension. If we perturb the 
initial conditions with larger values of perturbations, for example 
$\Delta x=-6 \times 10^{-2}$, we observe again two  tori with smooth color 
variation  for 1000 consequents as in the previous case. However, if we 
continue the integration of the orbits  we observe that the consequents  
start to deviate from two  tori and they form a cloud  in the 3D projection of 
the 4D surface of section around them (Fig.~\ref{2tor2}). In the cloud  the color is mixed and this means that the 
points are far away in the 4h dimension. The dynamical behavior according to
which an orbit stays close to an invariant torus for some time and then
diffuses in phase space is typical of sticky orbits
[Contopoulos and Harsoula 2008]. In Table 1 we give the values and the
direction of the perturbations for which  we observe tori. Just beyond this interval we encounter sticky orbits as the one in Fig.~\ref{2tor2}.

\subsubsection{Phase space structure close to~u}
Close to $s$, at Ej= $-4.33035$, we trace also the  simple unstable 2-periodic orbit $u$ (Fig.~\ref{2tor1}).
Our code finds it at initial conditions $(x_0,z_0,\dot{x_0},\dot{z_0})=0.67649546,0.16719438,0.78985004,0.41351876)$.
In its neighborhood we have found two types 
of dynamical behavior. We can see the first type if  we perturb  for example the 
initial conditions by 
$\Delta x=10^{-4}$. In this case we observe in 
Fig.~\ref{2proj1} a ``ribbon'' that connects the  points of the simple 
unstable 2-periodic orbit. If we  apply the method of color and rotation we 
observe in Fig.~\ref{2sos1} a smooth color variation from red to 
blue. This means that the ``ribbon'' is a 4D structure. After 5300 
intersections we observe in Fig.~\ref{2proj1a}, that the consequents 
leave the ``ribbon'' and they scatter in the phase space. This orbit is also sticky 
in the sense that remains close to a given phase space structure for some time and then diffuses in the 4D space of section.

The second type of dynamical behavior  can be seen in the neighborhood of the
simple  unstable 2-periodic orbit for Ej= $-4.33035$. If  we  add for example
a perturbation in the z-direction $\Delta z= 2 \times 10^{-3}$  in  the 
initial conditions of $u$ we observe a filamentary surface in the 3D projection 
of the 4D surface of section (Fig.~\ref{2proj2}). This filamentary structure 
connects the points of $u$  and forms four loops. The two of them surround  the 
two tori that are around the two points  of s. In the regions, close to the points
 of $u$ we have two self-intersections of the filamentary 
structure. In Fig.~\ref{2proj2} we observe two more  self-intersections indicated with 
arrows and the formation of 4 loops. The loops surround the periodic points $s$ and two more points representing a periodic orbit symmetric to $s$ with respect to the equatorial plane $z=0$, while the arrows represent an orbit symmetric to $u$.
 The dynamics in the neighborhood of this second set 
of periodic  orbits is similar to the one close to $s$ and u. Now
we apply the method of color and rotation in order to study the distribution 
of the consequents in the 4th dimension (Fig.~\ref{2sos2}). We observe, that we
have also in this case a smooth color variation from red to blue and this 
means that the  consequents are on the filamentary structure in the 4D space 
of section. For more than 4700 intersections the points deviate from this structure and 
form a cloud of points around it. Therefore in this case we encounter again the
phenomenon of stickiness. Table 2  gives the  range of the perturbation of the
initial conditions for which we find the ``ribbons'' or the ``filaments'' 
connecting the points of u, for $E_j=-4.33035$. For larger values of the  
perturbations we find clouds of points as the one we presented in 
Fig.~\ref{2proj1a}.

\subsection{4D spaces of section in the neighborhood of  7-periodic 
orbits}
The structures observed close to the 2-periodic orbits, have been found in the 
neighborhood  of every $m$-periodic orbit we have studied in our system. The 
results are qualitatively similar, but the filamentary structure that joins
the points of the simple unstable orbits becomes complex as $m$ increases.
Below we give one more example of the phase space structure close to a high multiplicity orbit, this time around
a 7-periodic one. 

\subsubsection{Phase space structure close to s1}
We apply a perturbation both in the $x$-direction and $z$-direction in the initial 
conditions of the stable 7-periodic orbit of the family $s1$,  
$\Delta x= 1.6 \times 10^{-2}$ and $\Delta z=-5 \times 10^{-3}$, for 
Ej= $-4.622377$. At this Jacobi constant we find $s1$ at initial conditions 
$(x_0,z_0,\dot{x_0},\dot{z_0})=(0.25992717,0.29991487,0,0)$.
The perturbed orbit has been 
integrated for $10^{4}$ consequents. In 
the $(x,\dot x,z)$ 3D projection of the 4D surface of section we observe seven 
tori surrounding the  points with the initial conditions of $s1$
(Fig.~\ref{7tor}). In this projection we  observe, that two 
of the tori intersect each other. To check their position in the 4th dimension 
we apply  the method of color and rotation (Fig. \ref{7tor}). 
By taking into account the values of all tori in the 4th dimension, $\dot z$, and scaling the colors according to [$\dot z_{min}, \dot z_{max}$]=[$-0.617, 0.617$], 
we find that to each small torus correspond one or mostly two primary colors. We 
observe that in the case of the two  tori that intersect each other in the  
3D projection $(x,\dot x,z)$ we have the meeting  of different colors at their 
intersection. This means  that this intersection does not exist in the 
4D space but only in the 3D  projection, as expected. Next we apply the 
method of color and rotation only to  one torus, which we call T1 and which is 
indicated with  an arrow in Fig. \ref{7tor}. In Fig.~\ref{7tor2} we observe a  
smooth color variation on its surface, which obviously corresponds to a different $\dot z$ range than that in Fig.~\ref{7tor}, and we see a  succession of colors 
from red to orange, to  yellow, to green, to light blue, to blue and finally  
to violet. Comparing this result with this found  in KP11 we realize that we 
have a morphology typical for rotational tori [Vrahatis et al 1997, KP11]. The 
only difference is that we do not observe in the present case the transition 
of the color sequence from the external to the internal surface of the torus, 
as observed in cases around  stable simple periodic orbits in KP11 
(compare with figure 11 in KP11). 
\par If we perturb the initial conditions of the periodic orbit of $s1$ by 
$\Delta x =10^{-2}$ we see in Fig.~\ref{7tor1a} seven tori without any
intersection of these tori in the 3D subspace. These tori have also a smooth 
color variation (Fig.~\ref{7tor1a}). In Table 3 we give  the range of the
perturbation, for which we find tori in the neighborhood of s1.   

\begin{figure}
\begin{center}
\includegraphics [scale =0.46]{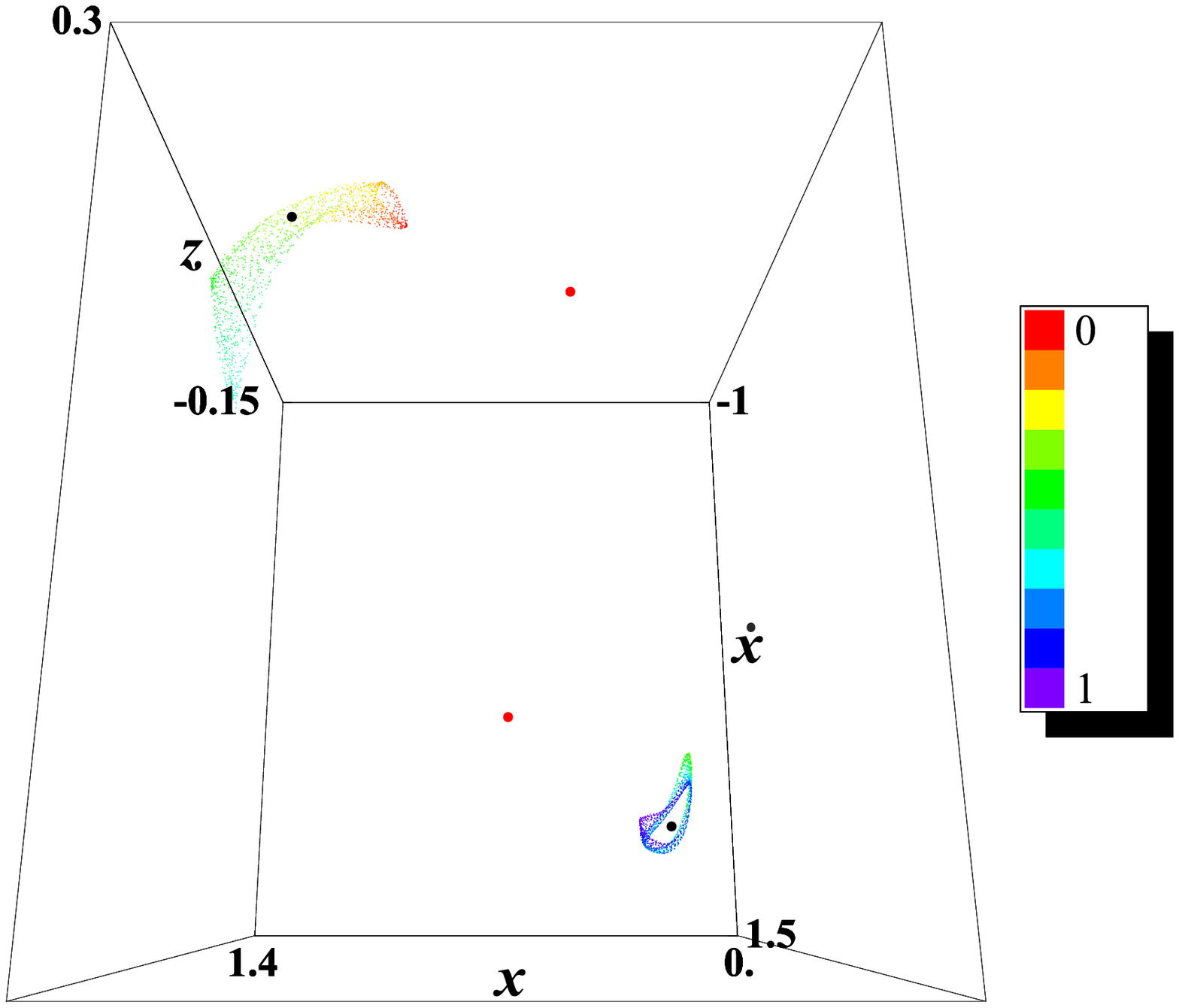}
\caption{The  4D surface of section for Ej=$-4.33035$ (3000 consequents) 
in the neighborhood of the stable 2-periodic orbit $s$. The initial conditions
of $s$ (black dots) are perturbed by $\Delta x= 4 \times 10^{-2}$. We use the 
$(x,\dot x, z)$ projection to depict the consequents and the $\dot z$  
value ($-0.253\leq \dot z \leq 0.708$) to  color them. We give also the initial conditions of the associated 
simple unstable  2-periodic  orbit $u$ with red dots. Our point of view in 
spherical  coordinates is $(\theta, \phi) = (22.5^{o},0^{o})$.} 
\label{2tor1} 
\end{center}
\end{figure}

\begin{figure}
\begin{center}
\includegraphics [scale =0.4]{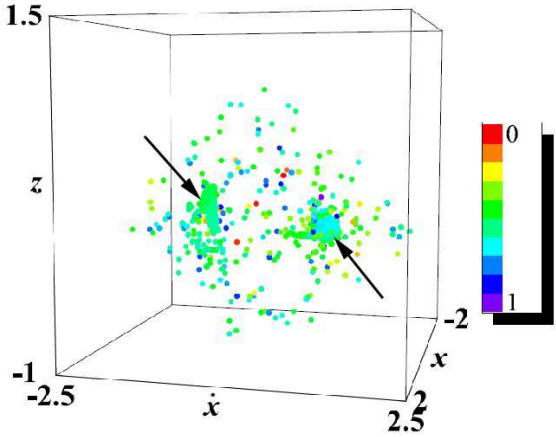}
\caption{1500 consequents of the sticky orbit we found in the 4D surface of section for Ej=$-4.33035$. Color is given to the consequents according to their $\dot z$ value ($-1.99\leq \dot z \leq 1.982$).
The orbit stays initially close to the region of the two tori (indicated with arrows) and then diffuses in phase space. We have chosen a $(\theta, \phi) = (22.5^{o},22.5^{o})$ point of view for a better inspection of the regions where the consequents stay initially concentrated and the way they diffuse.} 
\label{2tor2} 
\end{center}
\end{figure}

\begin{figure}
\begin{center}
\includegraphics [scale =0.7]{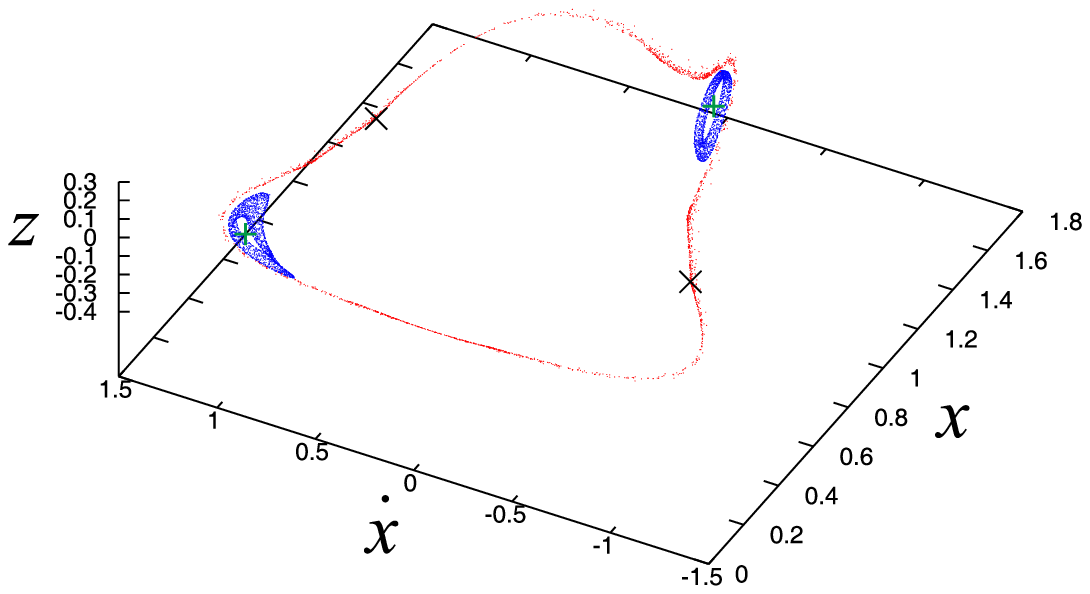}
\caption{ The 3D  subspace $(x,\dot x,z)$ of the 4D surface of section for 
Ej=$-4.33035$  in the neighborhood of $s$ and $u$. We apply a perturbation $\Delta x= 10^{-4}$ to their initial conditions and we consider 
5000 consequents. The initial conditions of $s$ are indicated  
with ``+'' in green color. We give also the initial conditions of 
$u$  with $\times$ symbols in  black.  Our point of 
view in spherical  coordinates is $(\theta, \phi) = (26^{o},298^{o})$. Around $s$ a double torus structure is formed, while the perturbed orbit in the neighborhood of $u$ forms a filamentary structure.}  
\label{2proj1}
\end{center}
\end{figure}

\begin{figure}
\begin{center}
\includegraphics[scale=0.48]{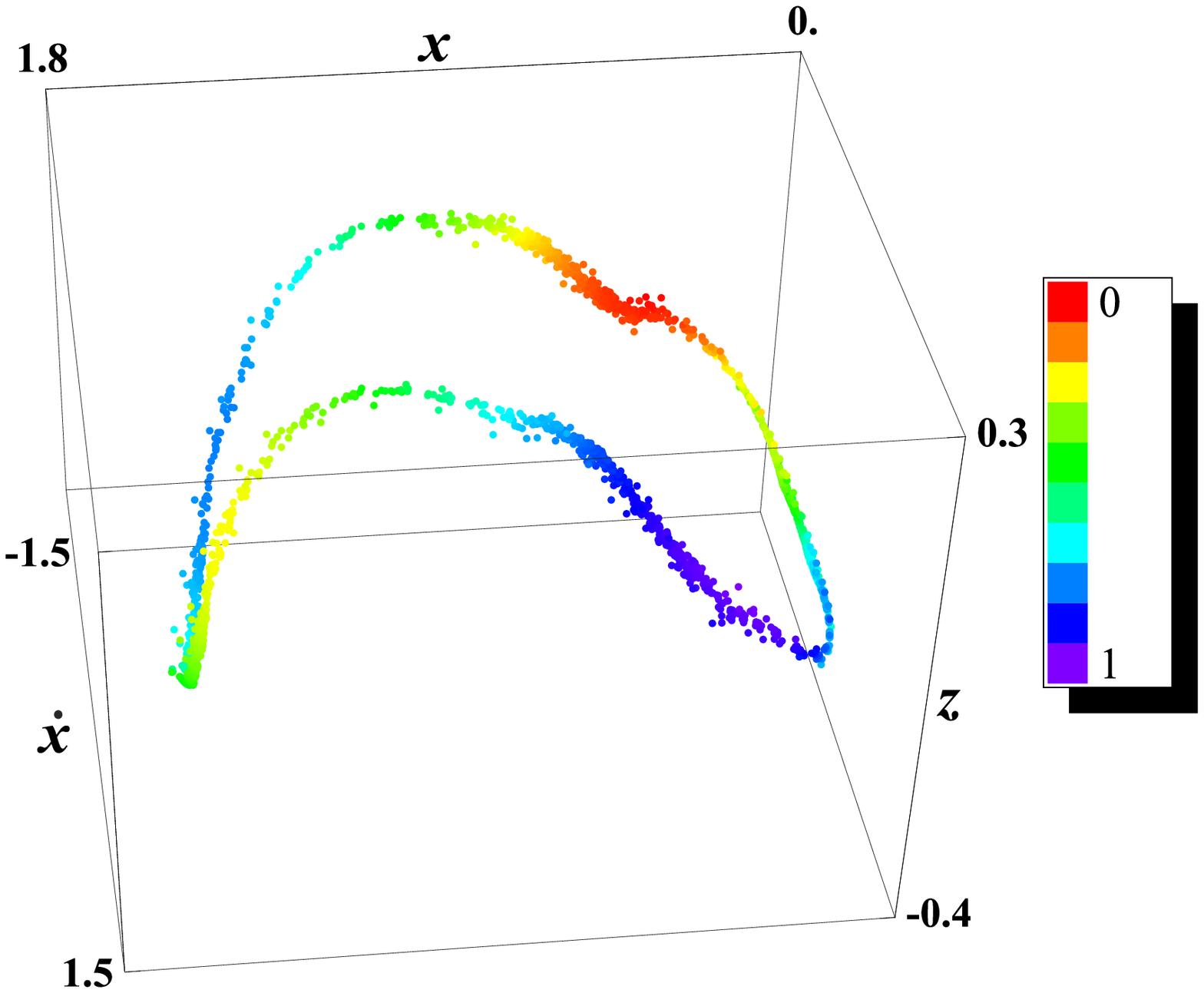}
\caption{The  orbital  behavior close  to  $u$ for a deviation $\Delta x= 10^{-4}$ 
from its initial conditions, at $E_j =-4.33035$ (5000 intersections)
in the 4D surface of section. We  
use  the  $(x,\dot x,z)$  space  for  plotting  the  points and  the  
$\dot z$  value  ($-0.637\leq \dot z \leq0.638$) to  color  them. Our point  of view  
in  spherical coordinates is  given  by  $(\theta, \phi) = (180^{o},22.5^{o})$.}
\label{2sos1}
\end{center}
\end{figure}

\begin{figure}
\begin{center}
\includegraphics[scale=0.78]{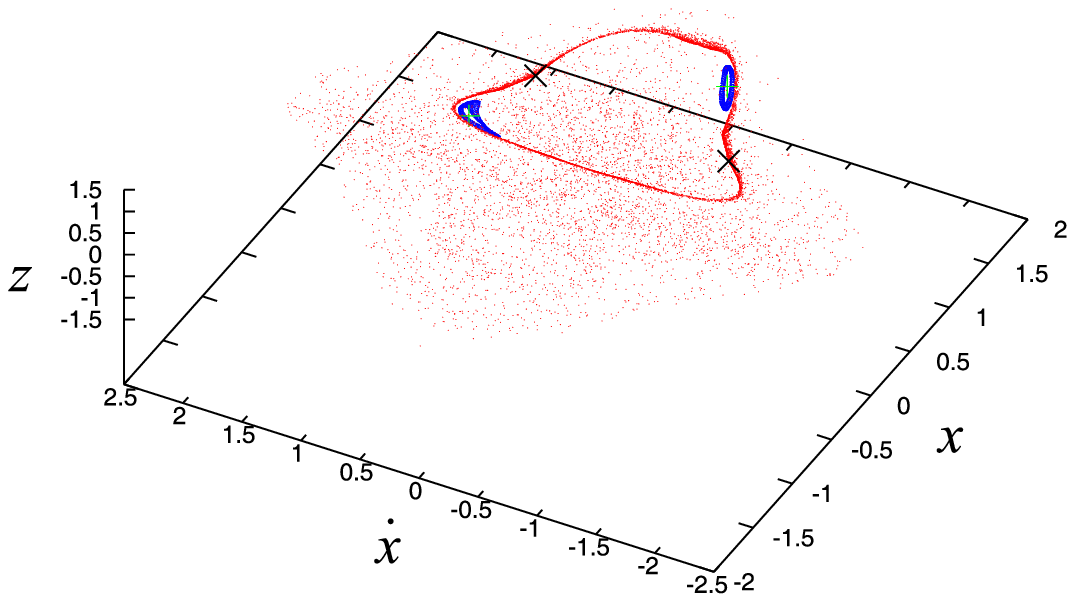}
\caption{A $(x,\dot x,z)$ 3D projection of the 4D surface of section close to the periodic orbits $s$ and $u$ ($\times$ symbol) for Ej=$-4.33035$. The ``ribbon'' we observe in red is the one given in 4D in Fig.~\ref{2sos1}, while the blue structures are the tori around $s$. This time we integrate the orbit of Fig.~\ref{2sos1} to get 7500 (instead of 5000) consequents. We observe a cloud of points, that surrounds both the tori and the ``ribbon''. Our point  of view  in  spherical coordinates is $(\theta, \phi) = (26^{o}, 240^{o})$.}
\label{2proj1a}
\end{center}
\end{figure}

\begin{figure}
\begin{center}
\includegraphics [scale =0.7]{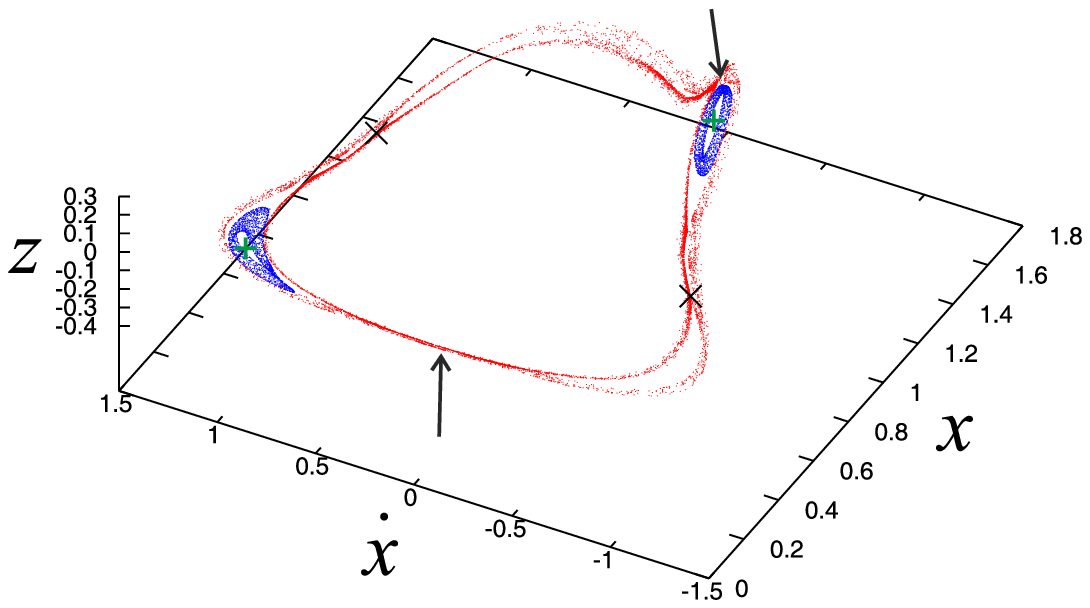}
\caption{The same projection as in Fig.~\ref{2proj1a}, but with an orbit close to $u$ perturbed this time in the $z$-direction by $\Delta z=2 \times 10^{-3}$
(4500 consequents). This time the ``ribbon'' has four self-intersections in the 3D subspace. We indicate $s$ with green ``+'' symbols, $u$  with ``$\times$'' in red, while the arrows point to the two additional self-intersections of the ``ribbon''. Our point of 
view is $(\theta, \phi) = (26^{o},298^{o})$.}  
\label{2proj2}
\end{center}
\end{figure}

\begin{figure}
\begin{center}
\includegraphics[scale=0.51]{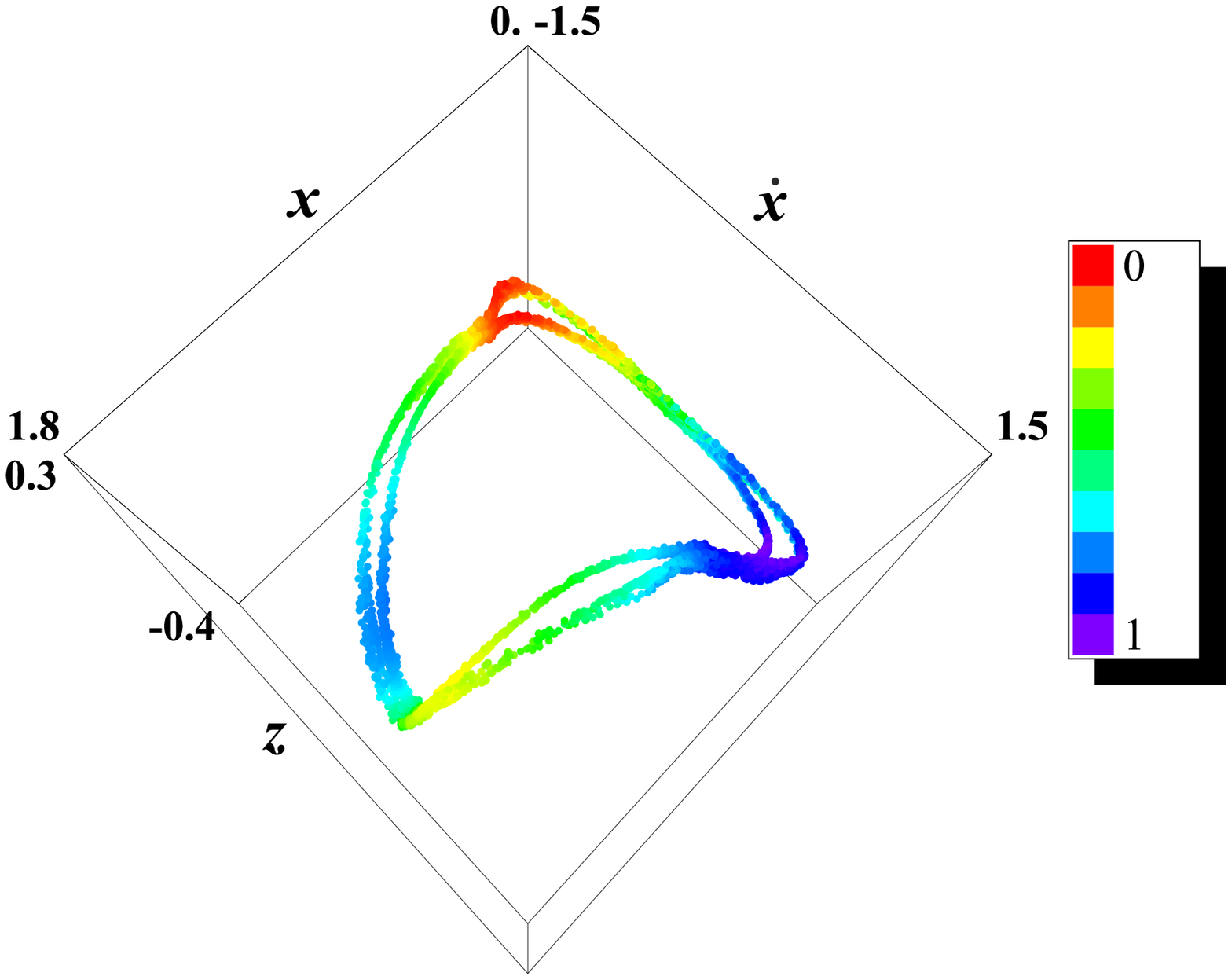}
\caption{The 4D representation of the ``ribbon'' of Fig.~\ref{2proj2}, for 4000 intersections. The consequents are colored according to their $\dot z$  value ($-0.651\leq \dot z \leq0.651$). The branches of the ``ribbon'' that intersect themselves, have the same color at the intersections, meaning that they are true intersetions in the 4D space. Our point  of view  
is  given  by  $(\theta, \phi) = (22.5^{o},22.5^{o})$.}
\label{2sos2}
\end{center}
\end{figure}

\begin{figure}
\begin{center}
\includegraphics [scale =0.4]{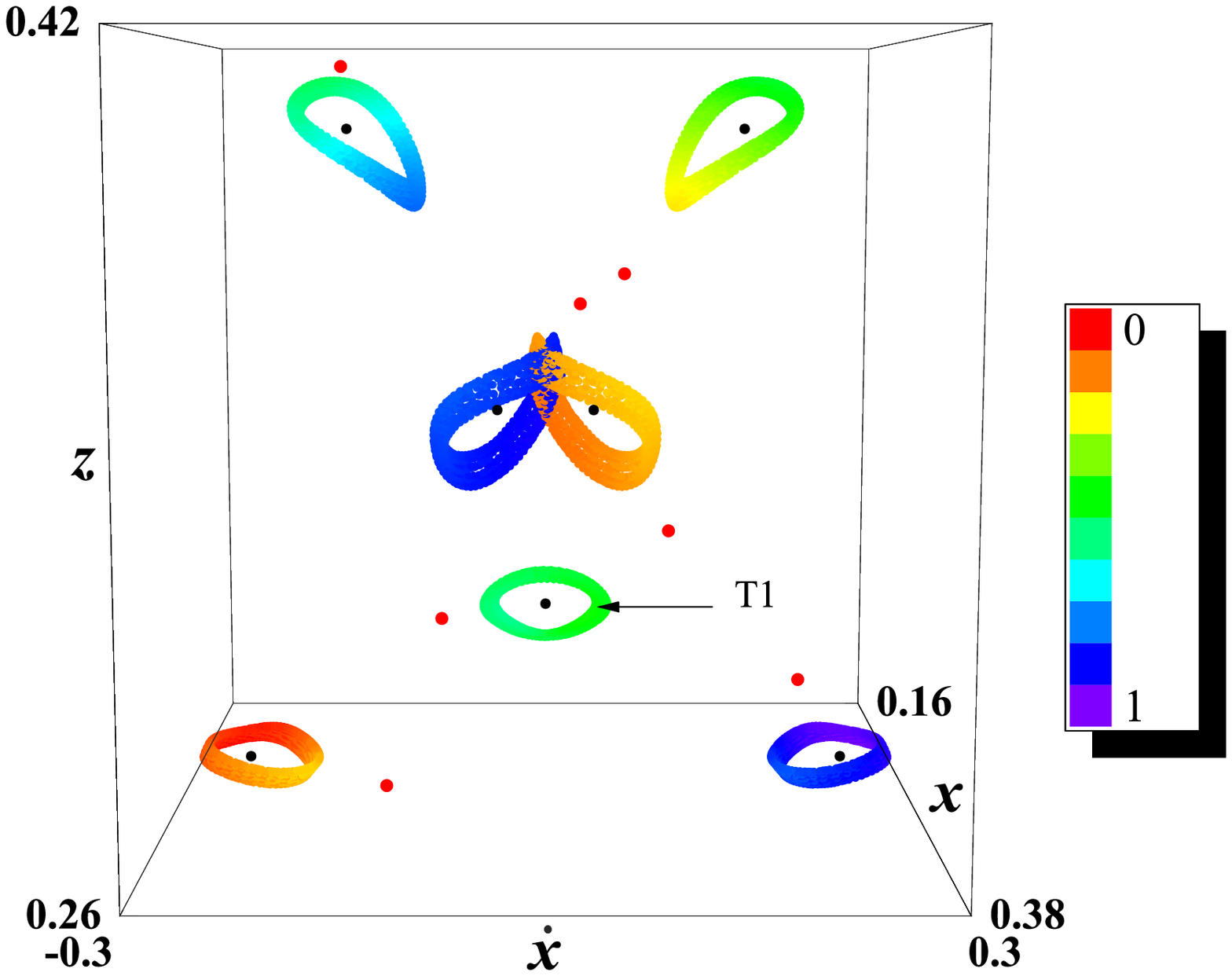}
\caption{The 4D surface of section for Ej=$-4.622377$ (6500 consequents)
in the neighborhood of $s1$. The initial conditions
of $s1$ are given with black dots. The orbit of the tori is found by perturbing 
the $s1$ initial conditions (see text) by $\Delta x= 1.6 \times 10^{-2}$ and  $\Delta z= -5 \times 10^{-3}$. The 
$(x,\dot x, z)$ projection is used to depict the consequents and the $\dot z$  
value ($-0.617\leq \dot z \leq0.617$) to color them. We give also the initial conditions of the associated 
simple  unstable  7-periodic  orbit $u1$ with  red dots. Our point of view in 
spherical  coordinates is $(\theta, \phi) = (0^{o},180^{o})$. We indicate with an 
arrow the T1 torus, which is discussed in the text. 
} 
\label{7tor} 
\end{center}
\end{figure}

\begin{figure}
\begin{center}
\hspace{-4mm}
\includegraphics[scale=0.46]{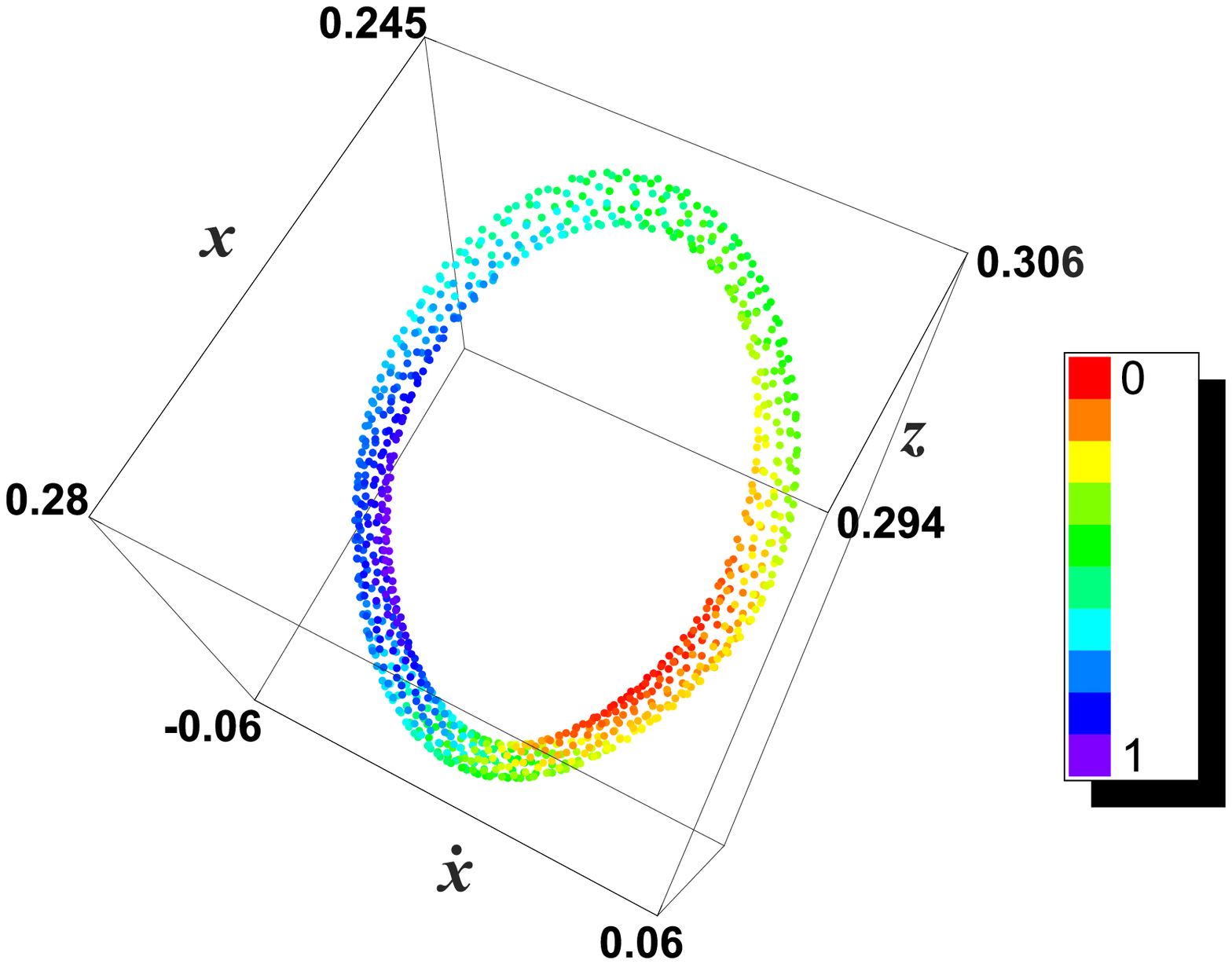}
\caption{Torus T1 in the 4D surface of section. Color is given to the consequents according to their $\dot z$ value ($-0.092\leq \dot z \leq0.092$). We observe a smooth color variation on its surface. Our  point  of  view  in  spherical coordinates is  given  by  $(\theta, \phi) = (22.5^{o}, 45^{o})$.}
\label{7tor2}
\end{center}
\end{figure}

\begin{figure}
\begin{center}
\hspace{-4mm}
\includegraphics[scale=0.45]{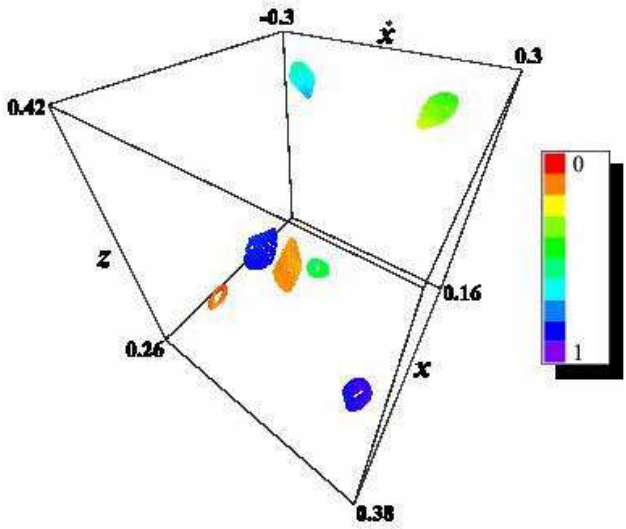}
\caption{Tori in the 4D surface of section for Ej=$-4.622377$ ($10^4$ 
consequents) in the neighborhood of the stable 7-periodic orbit $s1$  for 
$\Delta x= 10^{-2}$ . Color is given to the consequents according to their $\dot z$ value ($-0.592\leq \dot z \leq0.592$). Our  point  of  view  in  spherical coordinates is  
given   by $(\theta, \phi) = (22.5^{o}, 45^{o})$.}
\label{7tor1a}
\end{center}
\end{figure}

\begin{figure}
\begin{center}
\hspace{-20mm}
\includegraphics [scale =0.8]{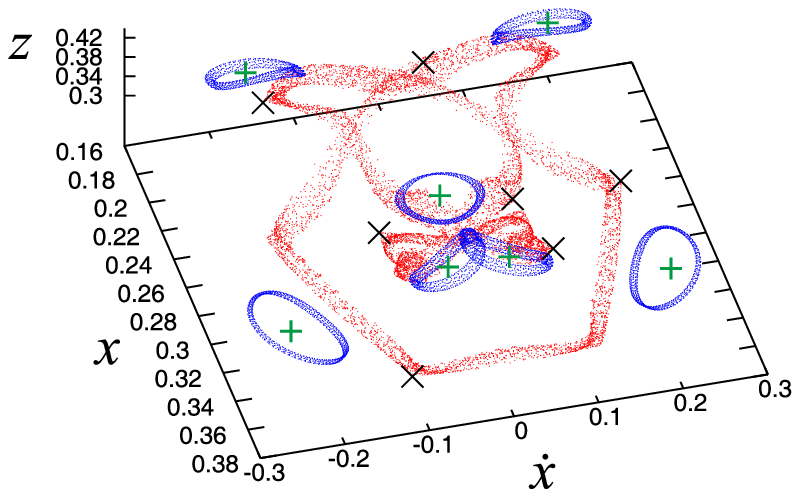}
\caption{The 3D $(x,\dot x,z)$ projection of the 4D surface of section for 
Ej=$-4.622377$  at the neighborhood of $s1$ and $u1$. Both $s1$, $u1$ initial conditions have been perturbed by $\Delta x= 10^{-4}$ to get a set of seven tori and a ``ribbon'' structure respectively. Green ``+'' correspond to $s1$ and black ``$\times$'' to $u1$. Our point of view in 
spherical coordinates is $(\theta, \phi) = (28^{o},48^{o})$.}  
\label{7proj1}
\end{center}
\end{figure}

\begin{figure}
\begin{center}
\includegraphics[scale=0.45]{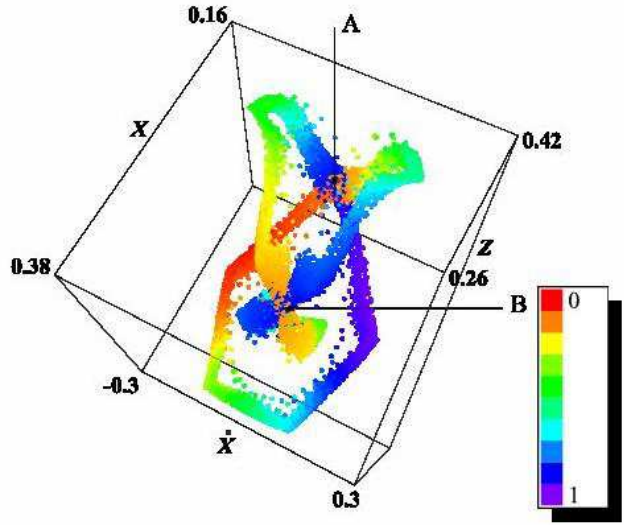}
\caption{The 4D surfaces of section in the neighborhood of $u1$ for the same orbit close to it depicted in
Fig.~\ref{7proj1}. We consider 6000 intersections.  We  
use  the  $(x,\dot x,z)$  space for  plotting  the  points and  the  
$\dot z$  value  ($-0.708\leq \dot z \leq0.693$) to  color  them. Different colors at the intersections (at ``A'' and ``B'') indicate that they are not true in 4D. Our point  of view  
is  $(\theta, \phi) = (22.5^{o},45^{o})$.}
\label{7sos1}
\end{center}
\end{figure}

\begin{figure}
\begin{center}
\hspace{-20mm}
\includegraphics [scale =0.7]{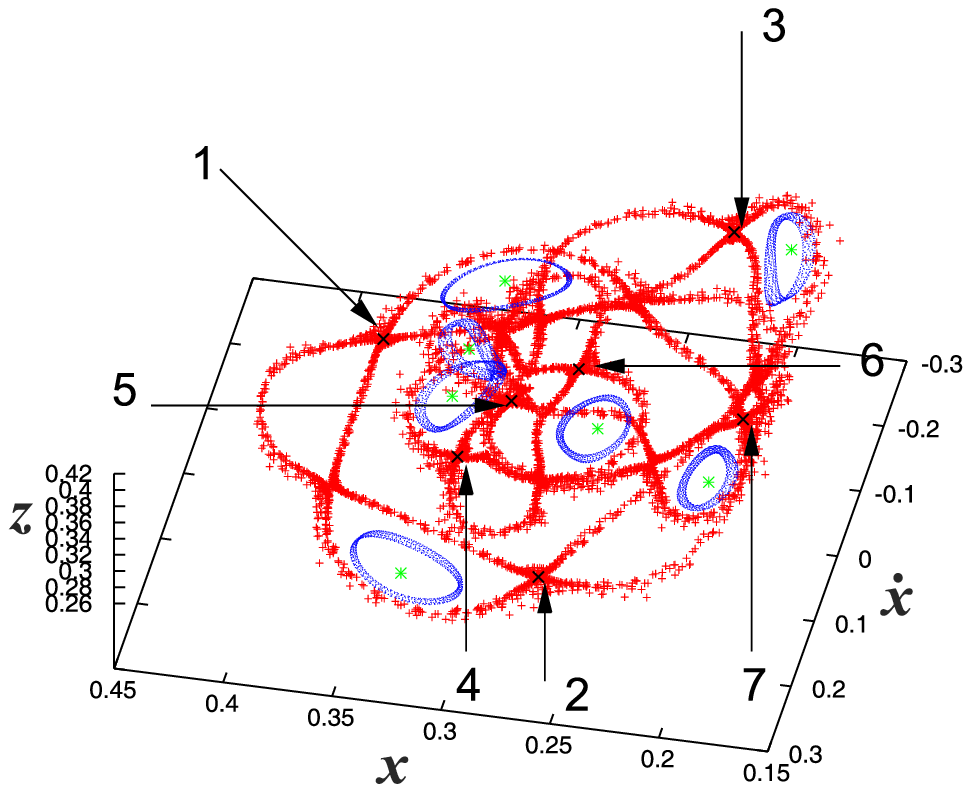}
\caption{ The 3D $(x,\dot x,z)$ projection of the 4D surface of section for 
Ej=$-4.622377$  at the neighborhoods of $s1$ and $u1$ for a perturbation of their initial conditions by $\Delta z=-3.6 \times 10^{-4}$. The initial conditions of $s1$ are indicated  with 
``$\times$'' and green color and those of $u1$  with ``$\times$'' black symbols. These points are 
numbered from 1 to 7. Our point of view in spherical  coordinates is 
$(\theta, \phi) = (26^{o},192^{o})$.}  
\label{7proj2}
\end{center}
\end{figure}

\begin{figure}
\begin{center}
\includegraphics[scale=0.51]{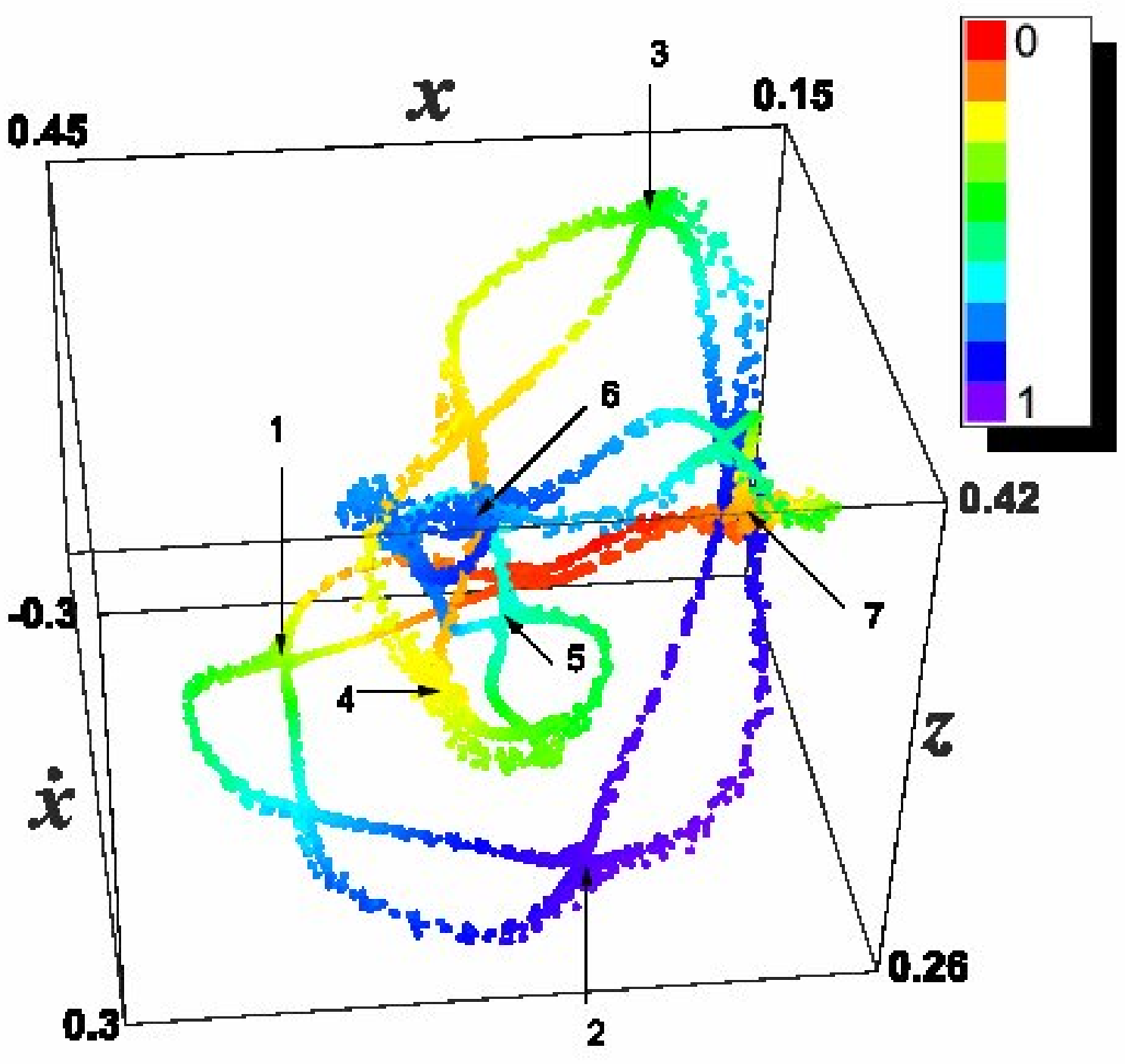}
\caption{The 4D representation of the orbit close to $u1$, which we give in Fig.~\ref{7proj2}, for 6000 intersections. We  use  the  $(x,\dot x,z)$  space  for  plotting  the  
points and  the  $\dot z$  value ($-0.726\leq \dot z \leq 0.709$) to  color  them. The points 1 to 7 indicate the $u1$ initial conditions. The common colors of the branches of the filamentary structures that meet at the intersections show, that they are true intersections in 4D. Our point  of view  
in  spherical coordinates is  given  by  $(\theta, \phi) = (180^{o}, 22.5^{o})$.}
\label{7stor}
\end{center}
\end{figure}

\subsubsection{Phase space structure close to u1}
Now we investigate the properties of orbits in the phase
space in the neighborhood of the simple unstable periodic orbit $u1$.
The orbit we study is at $E_j =-4.622377$ and is found with initial conditions 
$(x_0,z_0,\dot{x_0},\dot{z_0})=(0.27756296,,0.29935776,-0.084579242,0.16982513)$.
Firstly we add a perturbation in the initial conditions in 
the $x$-direction $\Delta x =10^{-4}$. In Fig.~\ref{7proj1} we see that the 
consequents that are depicted with red color double bow
in the 3D projection of the 4D surface of section that surrounds the 
seven tori that are around  the points of $s1$. If we 
color the consequents according to the value of their fourth dimension we see  
in Fig.~\ref{7sos1} that we have a smooth color variation from red to violet 
but at the two intersections A and B we have the meeting of different
colors (e.g. blue with orange). This means that we have different values of the
fourth dimension at the two intersections  and these intersections are only
projection effects and  not true intersections in the 4D space of section. The
consequents depart from the structure we give in Fig.~\ref{7sos1}, after 25000 
intersections.

\par If we perturb the same orbit by $\Delta z = -3.6 \times 10^{-4}$ 
at Ej= $-4.622377$. We observe in Fig. \ref{7proj2}, that the consequents  (with red) form  a filamentary  structure 
that  connects  the  points of the simple unstable  
7-periodic orbit and  surrounds the seven tori of the  stable 7-periodic 
orbit. In Fig. \ref{7proj2} with the numbers 1 until 7 we indicate the 7 points of $u1$ in the
$(x, \dot x, z)$ subspace. At these regions the filaments, that have been formed by evolving the orbit in time, cross each other in the 3D projection of the 4D surface of section 
(Fig. \ref{7proj2}). We observe, that besides the crossings at the numbered
regions, where we have the initial conditions of u1, we have 7 more crossings
of the filamentary structure by itself and the formation of 7 new loops, like
those surrounding the tori around the points of s1. We have again here, like
in the 2-periodic orbit in section 2.2.2, the presence of the symmetric
family, with respect to the equatorial plane.

In Fig. \ref{7stor} we observe a smooth color 
variation along the consequents that build the filaments. Starting from 1 we have a color 
succession  from green to light  blue, then to blue at  2, after that to light blue, then to 
green at 3, etc. At  the regions, where we have the self crossing of the filaments, i.e. close to the points 1, 2,...7, we observe that the regions are characterized by just one color only. For 
example at 1 the shade is green, at  2 blue etc. This  means that the fourth 
coordinate $\dot z$  of the consequents at these regions  has this time the 
same value  and  the  intersections  in  the 3D projections are real intersections  in the 
4D space. We also observe that the color evolves along both branches that 
depart from, or arrive at the points of the unstable periodic orbit. Both branches
show the same color evolution between two successive crossings.

The smooth succession of the colors along the filaments is observed as long as
the consequents participate in the filamentary structure. However, if we
consider more than about 6500 consequents in the neighborhood of the $u1$ orbit,
the points diffuse, occupying a large volume in the phase space and finally they
form a cloud in the 3D projections that surrounds the structure we observe in
Fig. \ref{7stor}. The dimensions of this cloud becomes about 15 times larger
in the x direction than the  structure depicted in Figs. \ref{7proj2} and
\ref{7stor}. This is determined by the space in which the particles are
allowed to move for this Ej. The cloud of points can be seen in
Fig.~\ref{cloud}. The red configuration in the central region of the diagram
is the structure we observe in Fig. \ref{7stor}. By giving colors to the
consequents according to their distribution in the 4th dimension we have
realized that in the cloud the distribution of points in the 4th dimension is
mixed (Fig. \ref{7sos2}). In Table 4 we summarize the range of perturbation
for which the consequents remain on the filamentary structure.

\begin{figure}
\begin{center}
\includegraphics[scale=0.85]{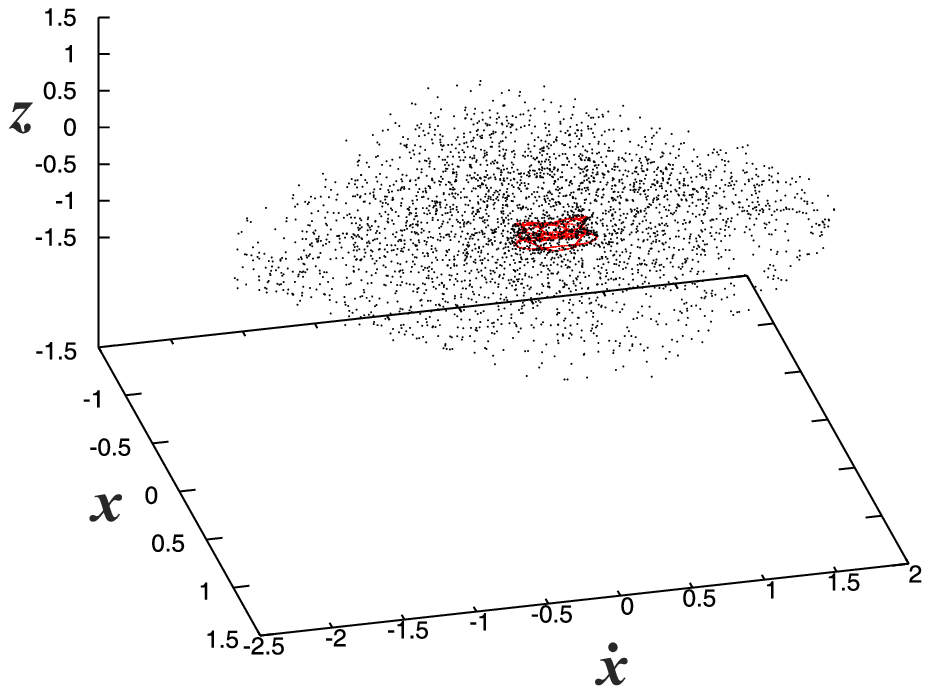}
\caption{The 3D $(x, z, \dot x)$ projection showing a cloud of points that surrounds the orbits close to $s1$ and $u1$ presented in Figs.~\ref{7proj2} and \ref{7stor}. This time we 
integrate the same orbit in the neighborhood of $u1$ as in the previous figures, 
but we consider 10000 instead of 6000 consequents. The red structure in the center of the cloud includes everything plotted in Figs.~\ref{7proj2}. Our point  of view  in  
spherical coordinates is  given by  $(\theta, \phi) = (37^{o}, 17^{o})$.}
\label{cloud}
\end{center}
\end{figure}

\begin{figure}
\begin{center}
\includegraphics[scale=0.8]{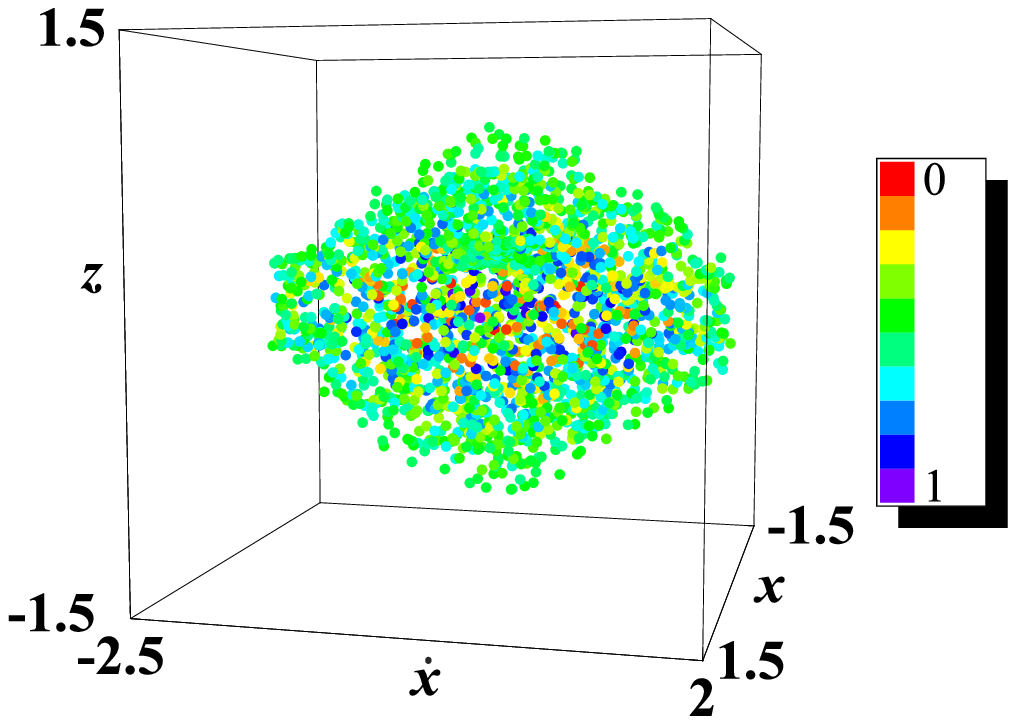}
\caption{The cloud of Fig.~\ref{cloud} in the 4D space of section. We plot 7500 consequents. We use the  $(x,\dot x,z)$ space for plotting  the points and  the  $\dot z$  value ($-2.29\leq \dot z \leq2.33$) to  color them. We observe mixing of colors. Our point of view is  $(\theta, \phi) = (180^{o}, 22.5^{o})$.}
\label{7sos2}
\end{center}
\end{figure}

\begin{table*}
\begin{center}
\tbl{The intervals of the perturbations of the initial conditions of the s
  orbit at $E_j =-4.33035$, for which we find 4D tori around s.}
{\begin{tabular}{|c|c|c|c|} \hline 
$\Delta x$&$\Delta \dot x$&$\Delta z$ &$\Delta \dot z$\\ \cline{1-4} 
from $10^{-4}$ to $4\times 10^{-2}$&from $10^{-4}$ to $3\times
10^{-2}$ &from $10^{-4}$ to $10^{-3}$&from $10^{-4}$ to
$6\times 10^{-2}$\\ \cline{1-4}
from $-10^{-4}$ to $-5\times 10^{-2}$&from $-10^{-4}$ to $-3\times
10^{-2}$ &from $-10^{-4}$ to $-10^{-3}$&from $-10^{-4}$ to
$-1.1\times 10^{-2}$\\ \hline
\end{tabular}}
\end{center}
\end{table*}

\begin{table*}
\begin{center}
\tbl{The range of perturbations in the initial conditions of $u$, that give
  ribbon or filamentary structure in the 4D surfaces of section  for 
$E_j =-4.33035$. The first two rows correspond to the  ribbons and the two 
last to the filaments.}
 {\begin{tabular}{|c|c|c|c|} \hline 
$\Delta x$&$\Delta \dot x$&$\Delta z$ &$\Delta \dot z$\\ \cline{1-4} 
from $10^{-4}$ to $3\times 10^{-3}$&from $10^{-4}$ to $2\times 10^{-2}$ &from 
$10^{-4}$ to $10^{-3}$&from $10^{-4}$ to $3 \times 10^{-3}$\\ \cline{1-4}
from $-10^{-4}$ to $-10^{-3}$&from $-10^{-4}$ to $-3\times
10^{-2}$&from $-10^{-4}$ to $-4\times 10^{-3}$& from $-10^{-4}$ to
$-9 \times 10^{-4}$\\ \cline{1-4}
from $4 \times 10^{-3}$ to $6 \times 10^{-2}$&from $3 \times
10^{-2}$  to $6 \times 10^{-2}$  &from $2 \times 10^{-3}$ to $10^{-2}$ &from
$4 \times 10^{-3}$ to $4 \times 10^{-2}$ \\ \cline{1-4}
from $-2 \times 10^{-3}$ to $-5\times 10^{-2}$&- &from $-5 \times
10^{-3}$ to $-2\times 10^{-2}$& from $-10^{-3}$ to $-3 \times 10^{-2}$\\ \hline
\end{tabular}}
\end{center}
\end{table*}

\begin{table*}
\begin{center}
\tbl{The range of perturbations in the initial conditions of s1, that give
  tori or tori with intersection in the 3D projections of the 4D surfaces of 
section for $E_j =-4.622377$. The first two rows correspond to the  tori  and the two last to the 
tori with intersection.}
 {\begin{tabular}{|c|c|c|c|} \hline 
 $\Delta x$&$\Delta \dot x$&$\Delta z$ &$\Delta \dot z$\\ \cline{1-4} 
from $10^{-4}$ to $1.3\times 10^{-2}$&from $10^{-4}$ to $3\times
10^{-2}$ &from $10^{-4}$ to $1.6\times 10^{-3}$&from $10^{-4}$ to
$10^{-1}$\\ \cline{1-4}
from $-10^{-4}$ to $-1.1\times 10^{-2}$&from $-10^{-4}$ to $-3\times
10^{-2}$&from $-10^{-4}$ to $-5\times 10^{-3}$& from $-10^{-4}$ to
$-10^{-1}$\\ \cline{1-4}
from $1.4 \times 10^{-2}$ to $1.8\times 10^{-2}$&from $4 \times
10^{-2}$  to $6 \times 10^{-2}$  &-& -\\ \cline{1-4}
from $-1.2 \times 10^{-2}$ to $-1.7\times 10^{-2}$&from $-4 \times
10^{-2}$ to $-6 \times 10^{-2}$  &-& -\\ \hline 
\end{tabular}}
\end{center}
\end{table*}

\begin{table*}
\begin{center}
\tbl{The range of perturbations in the initial conditions of $u1$, that give
  double bow or filamentary structure in the 4D surfaces of section  for 
$E_j =-4.622377$. The first two rows correspond to the double bows and the two 
last to the filaments.}
{\begin{tabular}{|c|c|c|c|} \hline 
 $\Delta x$&$\Delta \dot x$&$\Delta z$ &$\Delta \dot z$\\ \cline{1-4} 
from $10^{-4}$ to $7\times 10^{-4}$&from $10^{-4}$ to $3 \times
10^{-3}$ & from $10^{-4}$ to $2 \times 10^{-4}$&from $10^{-4}$ to $6 \times 10^{-3}$\\ \cline{1-4}
from $-10^{-4}$ to $-9 \times 10^{-4}$&from $-10^{-4}$ to $-4\times
10^{-3}$&from $-10^{-4}$ to $-2\times 10^{-4}$& from $-10^{-4}$ to
$-9 \times 10^{-3}$\\ \cline{1-4}
from $8 \times 10^{-4}$ to $1.6 \times 10^{-2}$&from $4 \times
10^{-3}$  to $2 \times 10^{-2}$  &from $3 \times 10^{-4}$ to $4 \times 10^{-3}$ &from
$7 \times 10^{-3}$ to $10^{-1}$ \\ \cline{1-4}
from $-10^{-3}$ to $-10^{-2}$&from $-5 \times 10^{-3}$ to $-8 \times
10^{-2}$ &from $-3 \times 10^{-4}$ to $-5\times 10^{-3}$& from $-10^{-2}$ to $-10^{-1}$\\ \hline 
\end{tabular}} 
\end{center}
\end{table*}

\section{Lyapunov Characteristic Numbers}
In this paper we study two kinds of orbits. The first kind is 
represented by tori in the 4D surface of section  in the neighborhood of 
stable 2-periodic and  7-periodic orbits. The second kind is represented 
initially  by structures confined in phase space (double bow, ribbon, ``filamentary'' structures etc.) 
and then by clouds in the 4D surfaces of section. These orbits are located in 
the neighborhood of simple unstable 2-periodic and 7-periodic orbits. In this 
section we calculate  the ``finite time''  Lyapunov Characteristic Number
(\textit{LCN}) for these two types of 
orbits.
\par The ``finite time'' Lyapunov Characteristic Number is defined as:
\begin{displaymath}
LCN(t)=\frac{1}{t}\ln\left|\frac{\xi(t)}{\xi(t_0)}\right|,
\end{displaymath}
where $\xi(t_0)$ and $\xi(t)$ are the distances between two points of two 
nearby orbits at times t = 0 and t respectively (see e.g. Skokos [2010])

\par Firstly, for the first kind of orbits,  we computed the maximal Lyapunov
Characteristic Number (\textit{mLCN}). For example, for the orbit that is represented 
in the 4D surface of section in Fig.~\ref{7tor1a}, the value of \textit{LCN(t)} decreases like 1/t and tends to zero as we can 
see in Fig.~\ref{7lcn}.

\par Then, we calculated the ``finite time'' Lyapunov Characteristic 
Number ($LCN(t)$) for the second type of orbits, for example for the orbit in Figs.~\ref{7stor} and 
Fig.~\ref{cloud}. In  Fig.~\ref{7stor} we have for the first 6500
consequents a filamentary structure. During this period the 
\textit{LCN(t)} of the orbit decreases to a value $1.4 \times 10^{-3}$
(Fig.~\ref{7lcn1}a). Beyond that point the orbit is represented by a cloud of
points in the 4D surface of section. During this phase, the  \textit{LCN(t})
fluctuates as  time increases and  finally increases and tends to level off 
around $1.67 \times 10^{-3}$ (Fig.~\ref{7lcn1}b).  
      
\begin{figure}[t]
\begin{center}
\begin{tabular}{cc}
\resizebox{80mm}{!}{\includegraphics{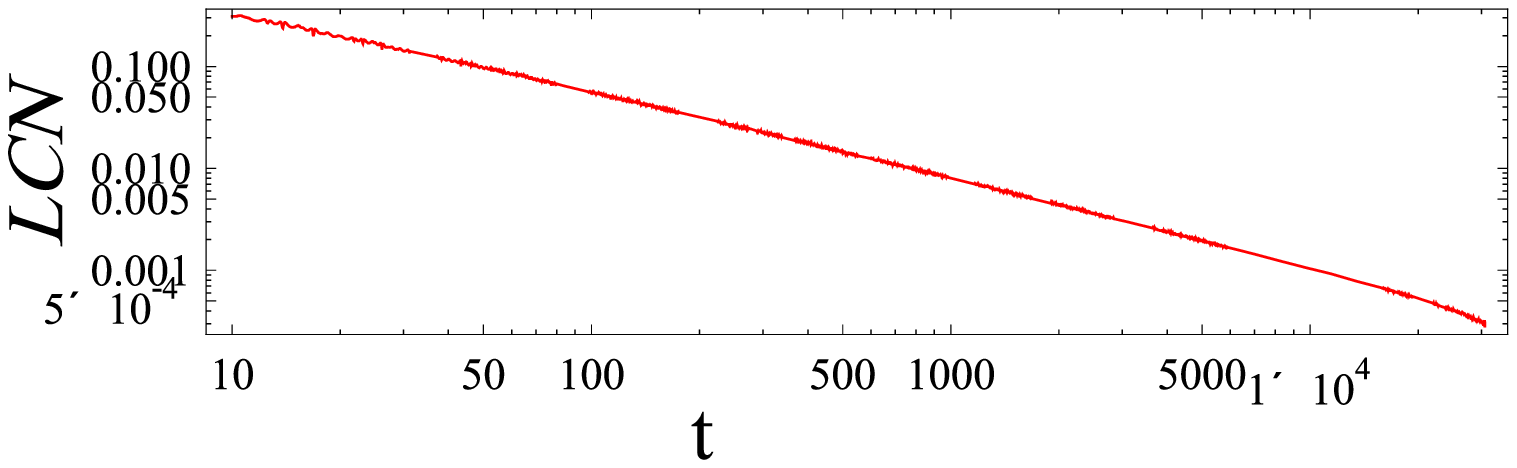}}\\
\end{tabular}
\caption{The \textit{LCN(t)} for the orbit given in the 4D surface of section 
in Fig.~\ref{7tor1a}. The axes are in logarithmic scale.}
\label{7lcn}
\end{center}
\end{figure}

\begin{figure}[t]
\begin{center}
\begin{tabular}{cc}
\resizebox{80mm}{!}{\includegraphics{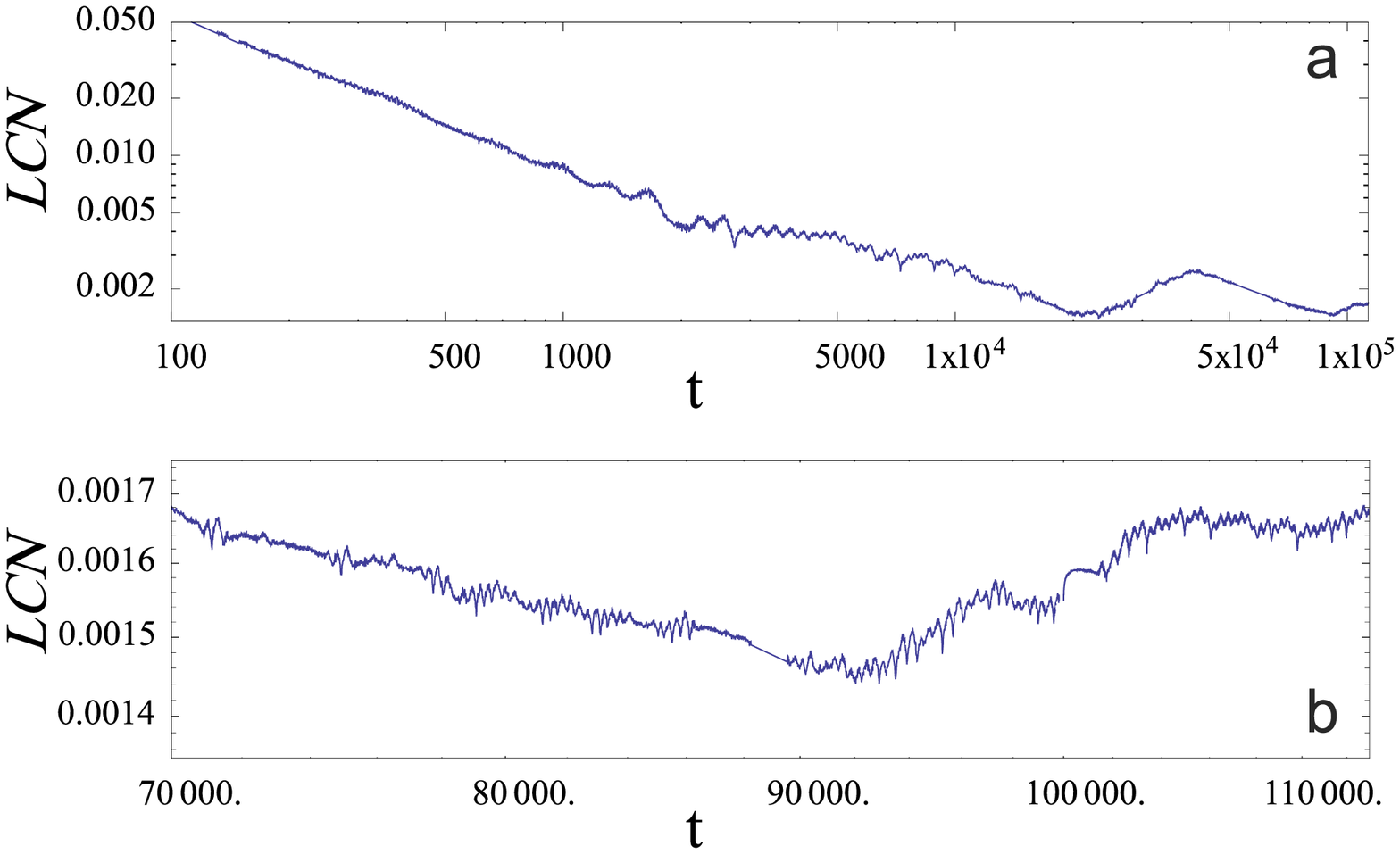}}\\
\end{tabular}
\caption{a)The evolution of  \textit{LCN(t)} for the orbit that is presented in Figs.~\ref{7stor} and Fig.~\ref{cloud}. (b) The part of Fig.~\ref{7lcn1}a for $70000\leq$t$\leq115000$.The axes are in logarithmic scale. }
\label{7lcn1}
\end{center}
\end{figure}

\section{Conclusions}
In this paper we studied the phase space structure in the neighborhood of  
stable and simple unstable periodic orbits in a rotating 3D galactic potential. We presented the dynamical behavior of two sets of stable-simple unstable orbits. The first was in the neighborhood of periodic orbits of multiplicity $m=2$, and the second in the neighborhood of periodic orbits of multiplicity $m=7$.  For less than 6000 
consequents we observed a direct correspondence between the standard
configuration, which we encounter in a resonance zone in 2D autonomous
Hamiltonian systems, and the structure of phase space in our 4D spaces of section, namely
a succession of elliptic and hyperbolic points. This means that also in
the case we study we observed a succession of stable and \textit{simple}
unstable points in the 4D spaces of section. For larger integration times the
consequents diffuse and occupy a larger volume of the phase space. We found
similar behavior for all cases of $m$-periodic orbits we studied in this system.

The following are the main conclusions from  our work: 
\begin{itemize}

\item In the neighborhood of the stable $m$-periodic orbit we found $m$ 
tori surrounding its initial conditions. We found smooth color variation along these tori in two ways. Along a particular torus and also along all $m$ tori, considering them as one object. This depends on the scale of the fourth coordinate, which gives the colors to the consequents. These tori are \textit{rotational} tori in the terminology introduced by Vrahatis et al. [1997]. 
Similar structures can be seen in the paper by Martinet \& Magnenat [1981].
However, we did not observe in the present study the color transition  from the external to the internal side of the torus on the individual tori as in some cases in KP11.
\item Integrating an orbit close to a simple unstable $m$-periodic orbit to obtain a few thousands of consequents, we found that they
  form a filamentary structure with smooth color variation in the 
  4th dimension. The filaments that are formed in this way connect the  points of the simple unstable $m$-periodic 
  orbit and surround the seven tori around the points of the stable 
  $m$-periodic orbit. The filamentary structures appear either as ``ribbons'' or as ``bows'' in the 4D spaces of section. 
\item In the regions close to the points of the simple unstable periodic orbit
  two branches of a filamentary structure meet and their consequents have
  the same color. This shows, that we have at these points self-intersections of the structures in the 4D space
  of section. 
\item The above described situation is a direct extrapolation of the typical case of a 2D autonomous Hamiltonian system with
  the chain of stability islands and the chaotic zone, which connects the points of
  the corresponding unstable periodic orbit. We note that in this 3D case the
  unstable orbit is \textit{simple} unstable and we have a smooth color variation along
  the filaments when we consider the first few thousands of the intersections in the space of section.
\item We encountered cases, where the consequents that remained on the filaments and reinforced this structure for a few thousand intersections, diffused later in the 4D space. The diffusion in the 4D space, was characterized by mixing of colors. 
\end{itemize}

\vspace{1cm}
\textit{Acknowledgments} We thank Prof. Contopoulos for fruitful discussions and valuable comments. MK is grateful to the ``Hellenic Center of Metals 
Research'' for its support in the frame of the current research.\\

\section{References}
Broucke R. [1969] ``Periodic orbits in the elliptic  restricted three-body 
  problem'' \textit{NASA Tech. Rep.} 32-1360, 1-125.\\
Contopoulos G. and Magnenat P. [1985] ``Simple three-dimensional  periodic 
  orbits in a  galactic-type potential'' \textit{Celest. Mech.} \textbf{37}, 
  387-414.\\ 
  Contopoulos G. [2002] \textit{Order and Chaos in  Dynamical Astronomy}
  Springer-Verlag, New York Berlin Heidelberg.\\
  Contopoulos G. and Harsoula M. [2008] ``Stickiness in Chaos''
  \textit{Int. J. Bif. Chaos} \textbf{18}, 2929-2949.\\
  Hadjidemetriou J.D. [1975] ``The stability of periodic orbits in the 
  three-body problem''\textit{Celest. Mech.} \textbf{ 12}, 255-276.\\
  Katsanikas M. and Patsis P.A. [2011] ``The structure of invariant tori in a
  3D galactic potential'' \textit{Int. J. Bif. Chaos} (in press) (KP11).\\
  Martinet L. and Magnenat P. [1981] `` Invariant surfaces and orbital 
  behavior in  dynamical systems with 3 degrees of freedom.'' \textit{Astron. 
    Astrophys.} \textbf{96}, 68-77.\\
  Patsis P.A. and Zachilas L. [1994] ``Using Color and rotation for 
  visualizing  four-dimensional Poincar\'{e} cross-sections:with applications 
  to the orbital  behavior of a three-dimensional Hamiltonian system'' 
  \textit{Int. J. Bif. Chaos}  \textbf{4}, 1399-1424.\\
  Skokos Ch. [2001] ``On the stability of periodic orbits of high dimensional
  autonomous Hamiltonian Systems'' \textit{Physica D} \textbf{159}, 155-179.\\
  Skokos Ch., Patsis P.A. and  Athanassoula E. [2002a] ``Orbital dynamics of 
  three-dimensional bars-I. The backbone of three-dimensional bars. A 
  fiducial  case'' \textit{Mon. Not. R. Astr. Soc.} \textbf{333}, 847-860.\\
  Skokos Ch., Patsis P.A. and Athanassoula E. [2002b] ``Orbital dynamics of 
  three-dimensional bars-II. Investigation of the parameter space''  
  \textit{Mon. Not. R. Astr. Soc.} \textbf {333}, 861-870.\\
  Skokos Ch. [2010]   ``The Lyapunov Characteristic Exponents and their 
  Computation'', \textit{Lect. Not. Phys.} \textbf{790}, 63-135.\\
 Vrahatis M.N., Isliker H. and Bountis T.C. [1997] ``Structure and breakdown
 of invariant tori in a 4-D mapping model of accelerator dynamics''
 \textit{Int. J. Bif. Chaos.} \textbf{7}, 2707-2722.\\

\end{document}